\documentclass[12pt,preprint]{aastex}

\def\apje{1}           
\ifodd\apje \usepackage{emulateapj5,epsfig} \fi

\def\omcen{$\omega$ Cen}
\def\OmCen{Omega Centauri}

\begin{document}

\shorttitle{\sc The Unique History of \omcen}
\shortauthors{\sc Gnedin et al.}

\title{The Unique History of the Globular Cluster $\omega$ Centauri}
     
\author{Oleg Y. Gnedin\altaffilmark{1}, 
        HongSheng Zhao\altaffilmark{2},
        J. E. Pringle\altaffilmark{1,2},
        S. Michael Fall\altaffilmark{1}, \\
        Mario Livio\altaffilmark{1}, and
        Georges Meylan\altaffilmark{1}}

\altaffiltext{1}{Space Telescope Science Institute, 3700 San Martin Drive,
  Baltimore, MD 21218; {\tt ognedin@stsci.edu}}
\altaffiltext{2}{Institute of Astronomy, Cambridge CB3 0HA, UK}

\begin{abstract}

Using current observational data and simple dynamical modeling, we
demonstrate that \omcen\ is not special among the Galactic globular
clusters in its ability to produce and retain the heavy elements
dispersed in the AGB phase of stellar evolution.  The multiple stellar
populations observed in \omcen\ cannot be explained if it had formed as
an isolated star cluster.  The formation within a progenitor galaxy of
the Milky Way is more likely, although the unique properties of
\omcen\ still remain a mystery.

\end{abstract}


\section{\OmCen, the most massive globular cluster in the Galaxy}

\OmCen\ is unique among globular clusters in the Milky Way in that its
member stars exhibit a wide spread in metallicity
\citep{FR:75,FN:81,MP:81}.  This has led to the general belief that
\omcen\ was chemically self-enriched, while the other clusters were
not (\citealt{SK:96}; see \citealt{MH:97} for a review).  In this
context, it is often noted that \omcen\ is the most massive globular
cluster in the Milky Way and thus possibly the most capable of
retaining its own stellar ejecta (e.g., \citealt{IA:00}).  The
globular cluster G1 in the halo of the Andromeda galaxy has similar
properties; it appears to be self-enriched and is also one of the most
massive clusters in its host galaxy \citep{Metal:01}.

\OmCen\ has several other intriguing and potentially related
properties.  The metal-rich stellar population is more centrally
concentrated and has a lower velocity dispersion than the metal-poor
population \citep{Netal:97}.  Moreover, the metal-rich population has
no detectable rotation, while the metal-poor population rotates
relatively rapidly ($\sim 7$ km s$^{-1}$), consistent with the
apparent flattening of the cluster, which is among the highest of all
globular clusters in the Galaxy \citep{FF:82}.  The metal-rich stellar
population also appears to be younger than the metal-poor population,
consistent with the idea of self-enrichment over a period $\Delta\tau
= 3-5$ Gyr (e.g., \citealt{HW:00,HR:00,Letal:99}).  This is supported
by the large enhancement in s-process elements in the metal-rich
population relative to that in globular clusters with similar
metallicities \citep{NDC:95,VWB:94}.  These properties have led to the
suggestion that \omcen\ managed to retain the ejecta from the first
generation of stars in the asymptotic giant branch (AGB) phase, which
then formed a centrally concentrated, metal-enriched, younger
generation of stars (e.g. \citealp{Netal:96,Setal:00}).

In this paper we demonstrate that \omcen\ is not special in the
ability to retain its own stellar ejecta.  The escape speed in \omcen\
is not the highest among the Galactic globular clusters.  We find that
more than half of the clusters could retain at least 20\% of their AGB
winds.  It is therefore surprising that \omcen\ is the only globular
cluster in the Milky Way to show signs of
self-enrichment.%
\footnote{
  \citet{SL:95} have inferred a dispersion of metallicity of 0.16 dex in
  M54, a globular cluster projected in the center of the Sagittarius
  dwarf galaxy.  Also, \citet{Retal:01} have suggested that two other
  massive globular clusters (NGC 6388 and NGC 6441) may have an age
  and/or metallicity spread of 1.2 Gyr and 0.15 dex, respectively.
  These estimates, however, are not as large as those for \omcen\ and
  may result from differential reddening or other systematic errors.
  Until these are confirmed, we assume that \omcen\ is the only Galactic
  globular cluster with multiple stellar populations.
}
We conclude that this cluster could not have evolved in isolation on
its present orbit in the Galaxy and that it must have been
substantially different in the past.  We discuss other models
suggested previously, including the idea that \omcen\ is the nucleus
of a dwarf galaxy that lost its outer envelop by tidal stripping
\citep{Zetal:88,F:93}.



\section{The escape velocity from the cluster}

To compute the escape velocities of Galactic globular clusters, we use
the photometric data from the catalog by \citet{H:96}.  Out of 147
clusters in the catalog, we choose 141 for which the data are
available on the distance, integrated magnitude, core radius, and
concentration parameter.  All clusters are fit by single-mass
isotropic King models with a constant mass-to-light ratio, $M/L_V = 3$
in solar units.

First, we check that the photometric models are consistent with the
directly measured central velocity dispersions for a sample of 56
clusters from \citet{PM:93}, their Table 2.  The most reliable
estimates of the dispersion are based on individual stellar radial
velocities, but these estimates depend on the position of stars
relative to the cluster center.  The core velocity dispersion of
\omcen\ is 17 km s$^{-1}$ \citep{MMM:97}, with the indication that it
rises in the center up to 25 km s$^{-1}$ \citep{vLetal:00}.

We thus avoid a direct comparison and show instead in Figure 1 two
separate histograms of the central velocity dispersions as inferred
from the photometric models and as measured directly.  The
distributions are qualitatively similar.  Also, the photometric model
of \omcen\ predicts a central velocity dispersion of about 17 km
s$^{-1}$, the same as the observed value.

\def\cap1{
  Distribution of the central line-of-sight velocity dispersion
  of the Galactic globular clusters, as inferred from the photometric models
  (left panel) and observed spectroscopically (right panel).  The arrows
  point to the velocity dispersion of \omcen.
}
\ifodd\apje 
  \centerline{\epsfig{file=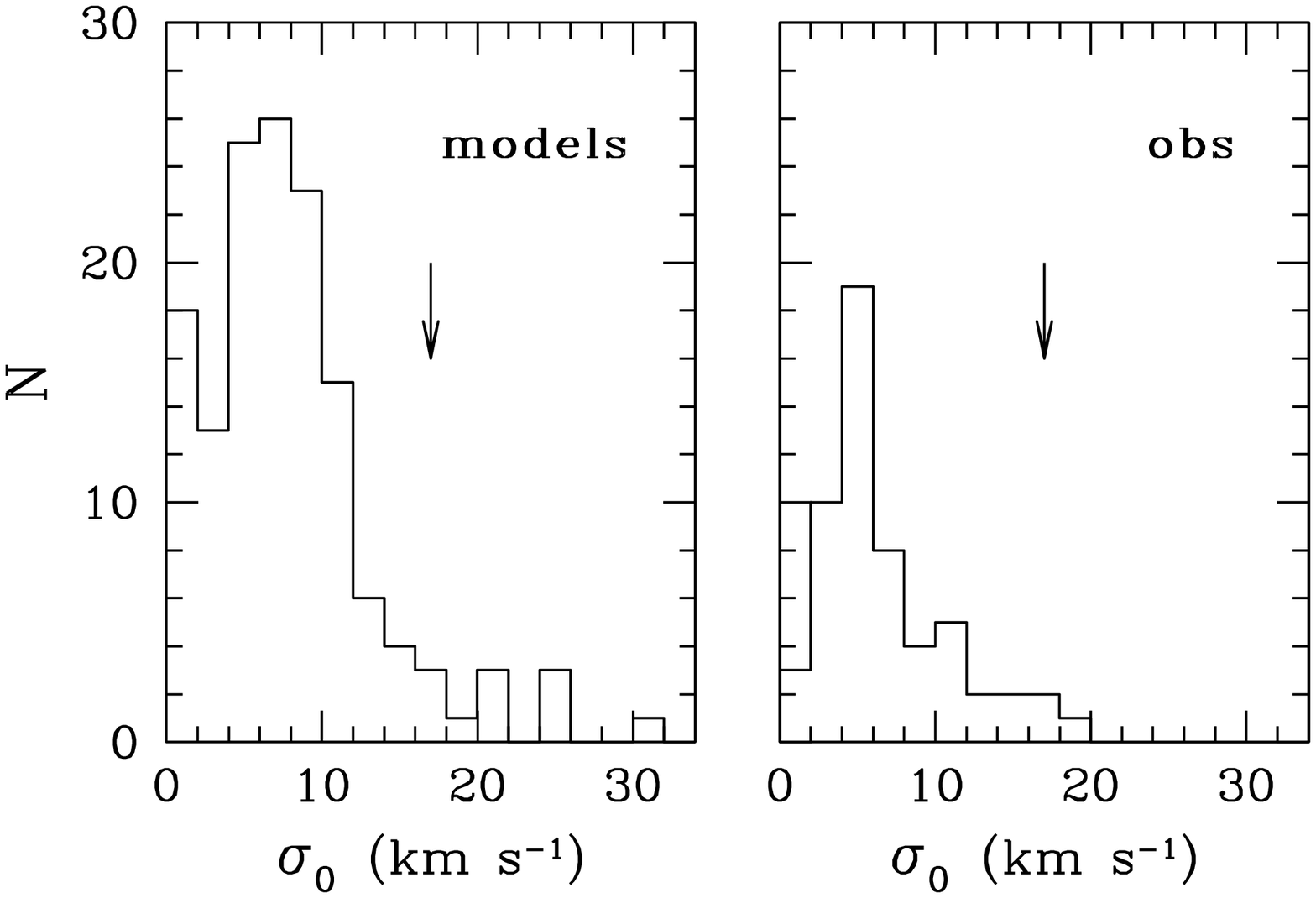,height=8.2cm}}
  \hspace*{0.5cm} Fig. 1.---{\small \cap1}
  \vspace*{0.5cm}
\else
  \begin{figure}
  \plotone{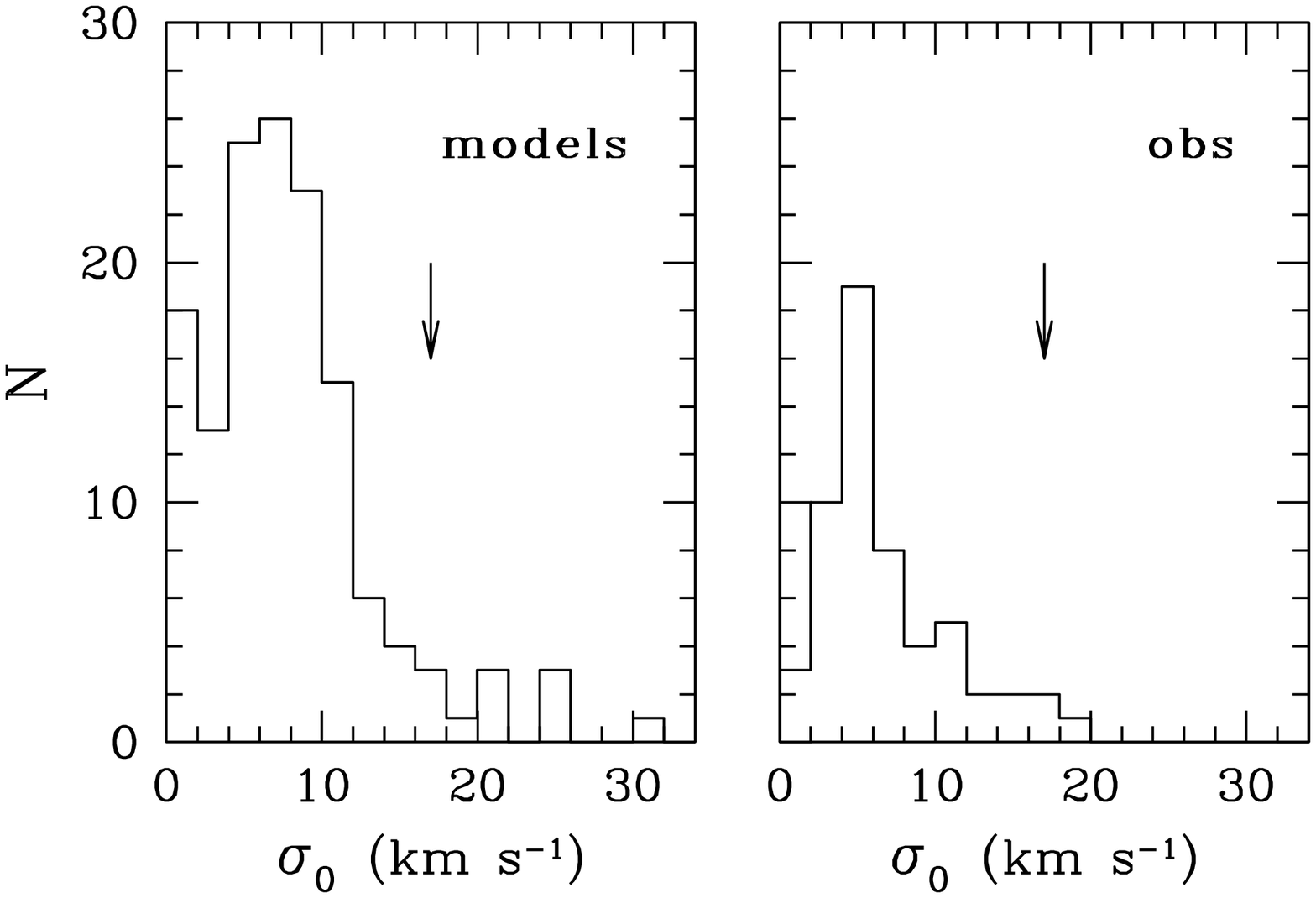} \label{fig:histo}
  \caption{\cap1}
  \end{figure}
\fi

We note that in the past 20 years there has been only one study
\citep{DMM:97} providing homogeneous velocity dispersion measurements
for a significant number of clusters.  Now, with the availability of
large-field survey spectrographs such data should be easier to obtain
and would provide a very useful constraint on the mass models of
globular clusters.

Having calibrated the photometric models, we can use them to determine
the escape velocity.  This is calculated using the reduced
gravitational potential, $v_{\rm esc} \equiv (2\Phi_t -2\Phi)^{1/2}$,
in the center (giving the maximum escape velocity) and at the
half-mass radius (providing a threshold for most stars).  Here, the
tidal potential $\Phi_t$ accounts for truncation at the tidal radius.

Figures 2 and 3 demonstrate that \omcen\ is not special in its escape
velocity.  There are 11 clusters with higher $v_{\rm esc,0}$ and 8
clusters with higher $v_{\rm esc,h}$.

We also find that the ratio of the central escape velocity to the
central velocity dispersion can be fit by a simple relation
\begin{equation}
  v_{\rm esc,0}/\sigma_0 = 3.7 + 0.9 \, (c-1.4),
\end{equation}
where $c$ is the concentration of the King models.  The fit is
accurate to 3\% in the range $c = 0.4 - 2.8$.

We now calculate the fraction of AGB winds retained by a cluster as a
function of the wind velocity $u_w$.  We integrate the distribution
function of stars, $f(E)$, over the phase space volume within which
the total velocity of the wind particles is below the escape speed at
the star's position:
\begin{equation}
  f_w = {1\over M} \int f(E) d{\bf r} d{\bf v}
        {1 \over 4\pi} \int d\Omega_u \,
        h(v_{\rm esc}(r) - \left| {\bf v} + {\bf u}_w \right|),
\end{equation}
where $\Omega_u$ is a solid angle of the wind velocity ${\bf u}_w$,
and $h(x)$ is a unit step function.

\def\cap2{
  Central escape velocity for 141 Galactic globular clusters
  vs their mass $M$.  \OmCen\ is denoted by an asterisk.  The escape
  velocity is calculated from the photometric single-mass King models
  with $M/L_V = 3$.
}
\ifodd\apje
  \centerline{\epsfig{file=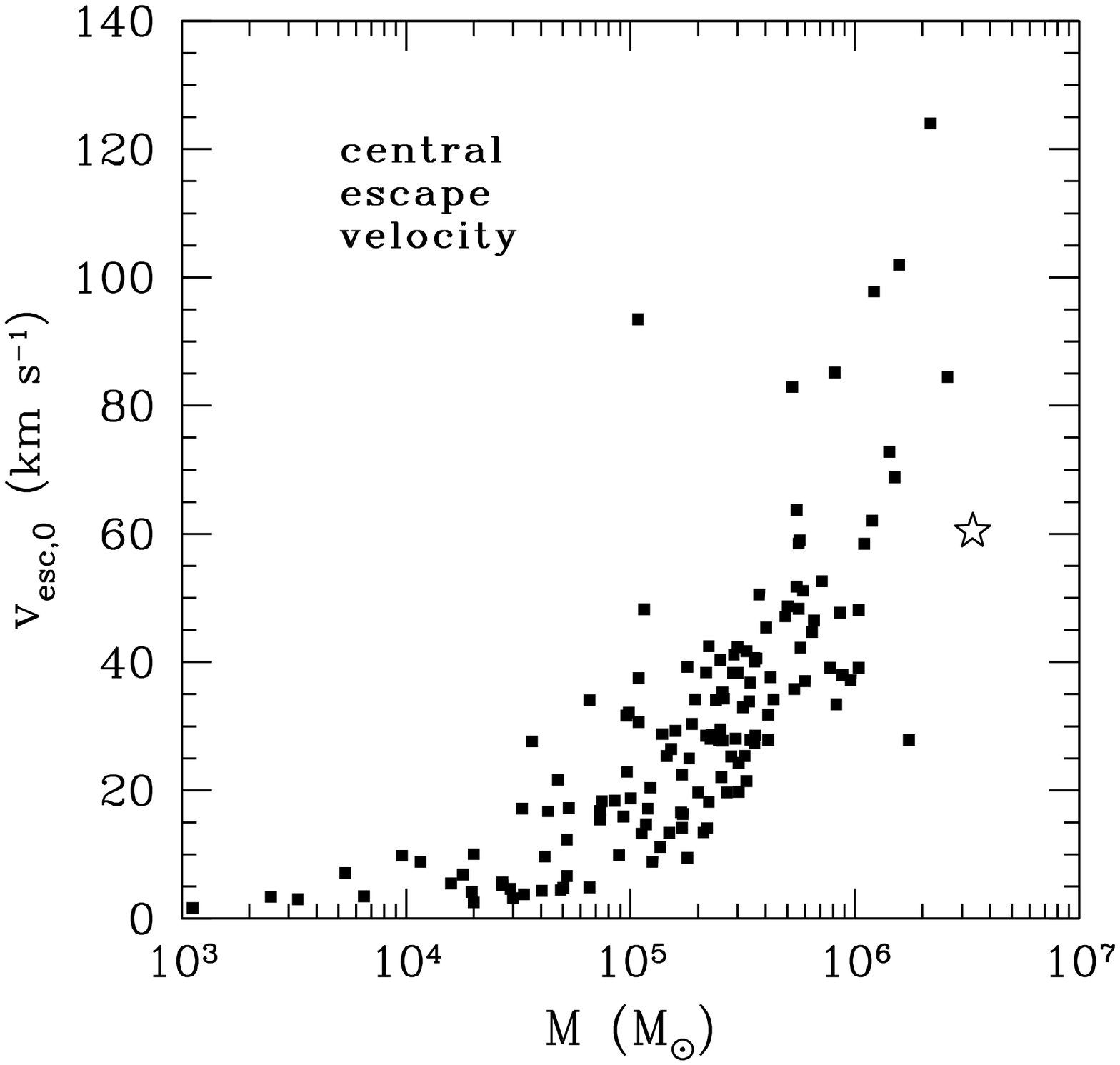,height=8.2cm}}
  \hspace*{0.5cm} Fig. 2.---{\small \cap2}
  \vspace*{0.5cm}
\else
  \begin{figure}
  \plotone{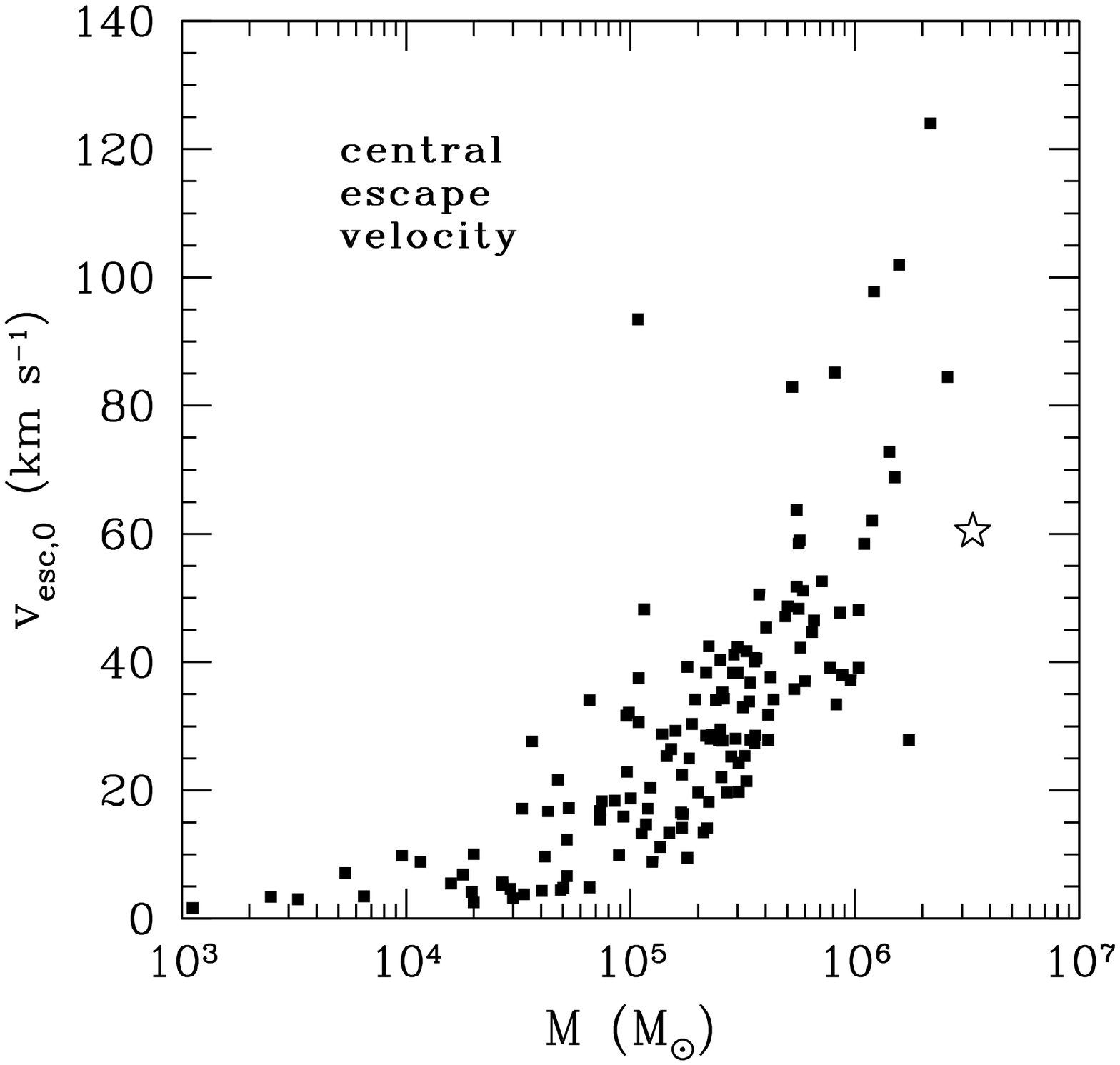} \label{fig:vesc0}
  \caption{\cap2}
  \end{figure}
\fi

\def\cap3{
  Escape velocity at the half-mass radius for the Galactic
  globular clusters vs their mass $M$.  \OmCen\ is denoted by an
  asterisk.
}
\ifodd\apje
  \centerline{\epsfig{file=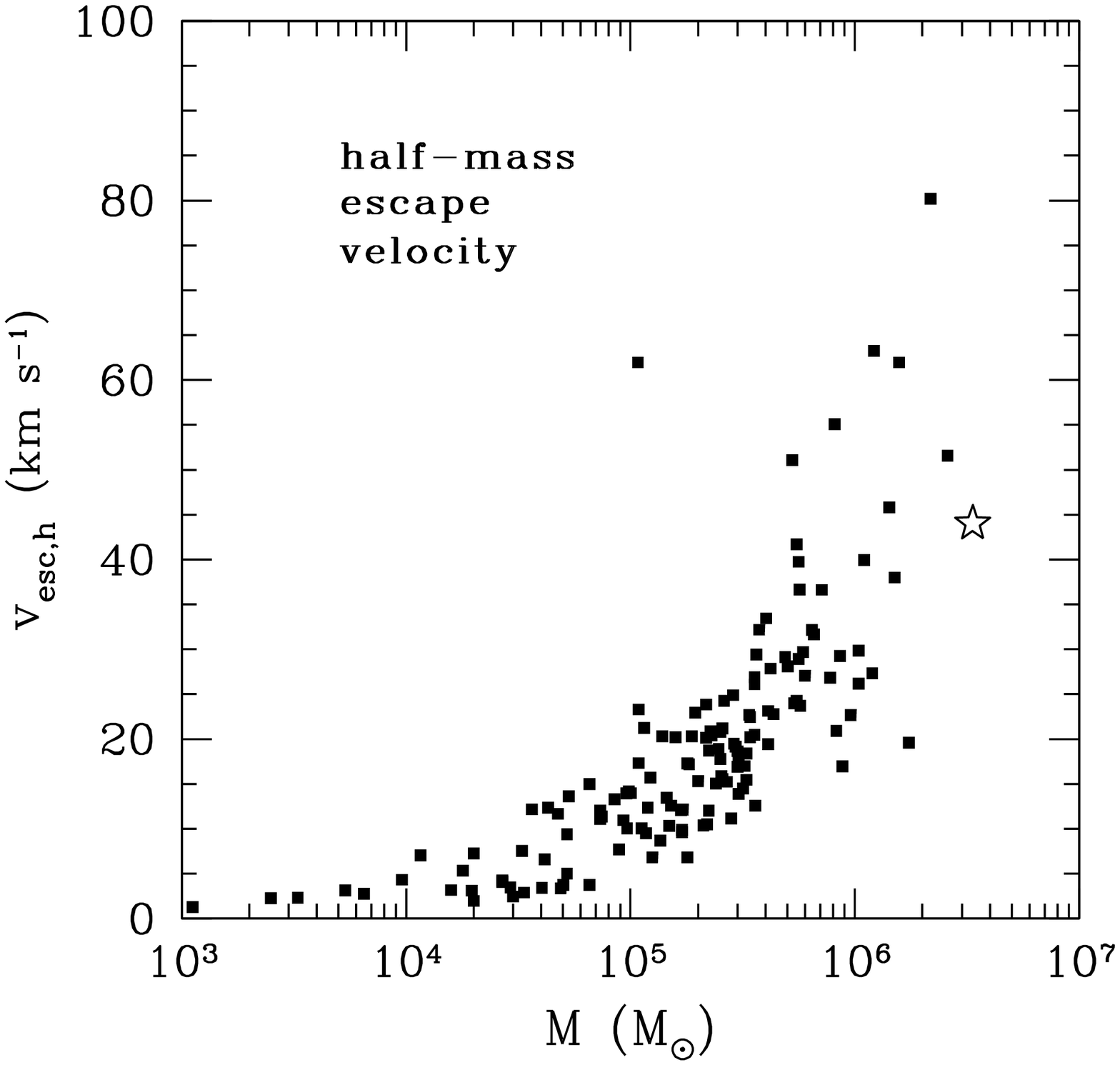,height=8.2cm}}
  \hspace*{0.5cm} Fig. 3.---{\small \cap3}
  \vspace*{0.5cm}
\else
  \begin{figure}
  \plotone{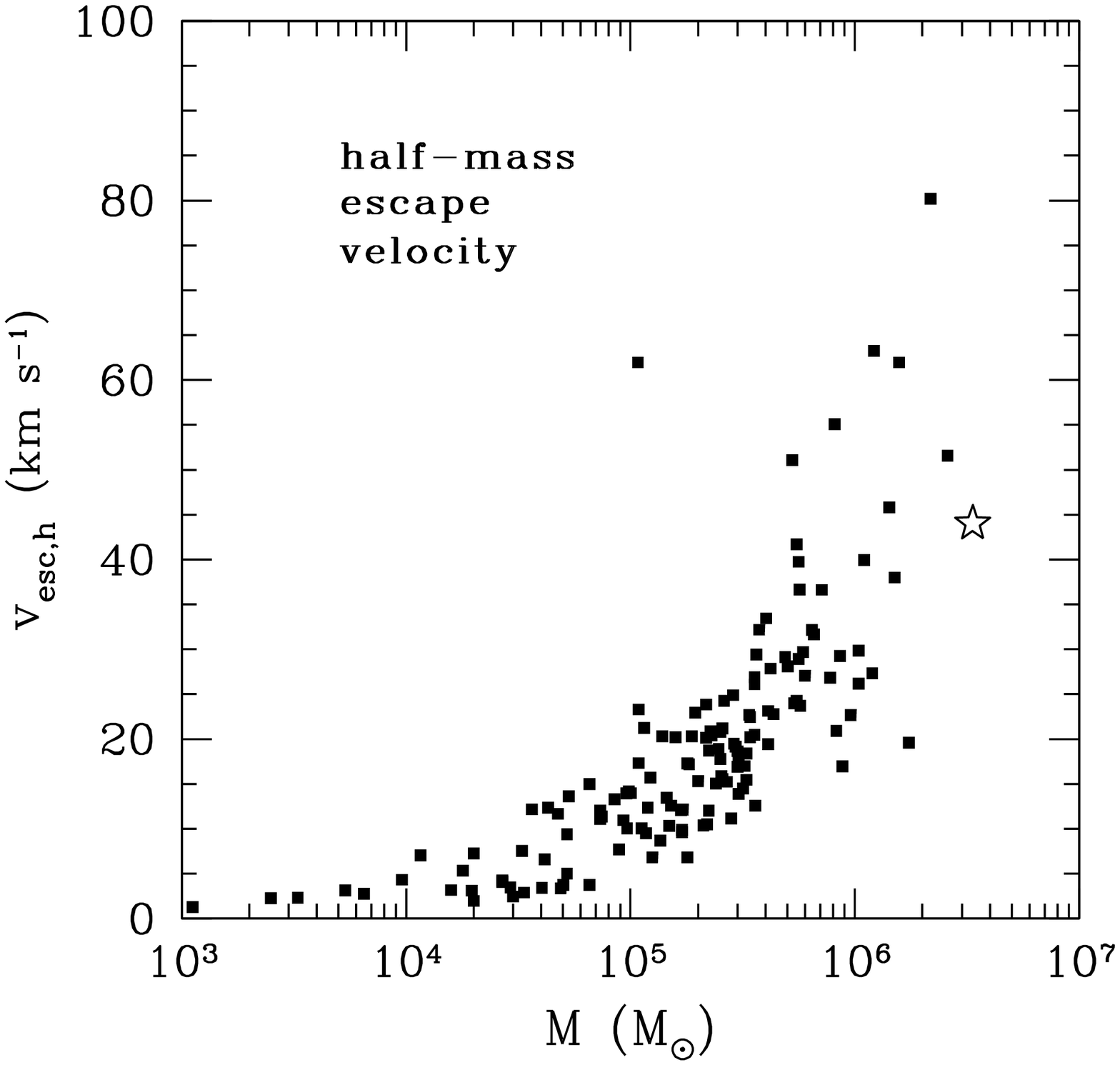} \label{fig:vesch}
  \caption{\cap3}
  \end{figure}
\fi

Figure 4 shows the result for clusters of various concentration
parameters, from $c=0.6$ to 2.4.  The fraction $f_w$ declines
gradually and even the clusters with $u_w \sim 3\sigma_0$ can retain
10 to 20\% of their winds.  Since according to \citet{Letal:93} the
characteristic terminal velocity of AGB winds is $u_w \approx 15$ km
s$^{-1}$ (although a velocity as high as 90 km s$^{-1}$ has been
measured for a field red giant by \citealt{DSL:92}), more than half of
the Galactic clusters should be able to retain a sizable fraction of
their winds.

\def\cap4{
  Fraction of AGB winds retained by clusters of concentration
  $c$ (marked by numbers) as a function of the wind velocity in units of
  the central velocity dispersion.  The top solid line shows the model
  with the highest binding energy ($c=1.4$), the dotted ($c=0.6$) and
  dashed ($c=2.4$) curves give the lower bounds.  All models with
  intermediate concentrations fall between the plotted lines.  For all
  concentrations, the fraction can be fit reasonably accurately by $f_w(x
  \equiv u_w/\sigma_0) = 1/[1 + (x/2)^2 + (x/4)^4 + (x/3)^8]$.
}
\ifodd\apje
  \centerline{\epsfig{file=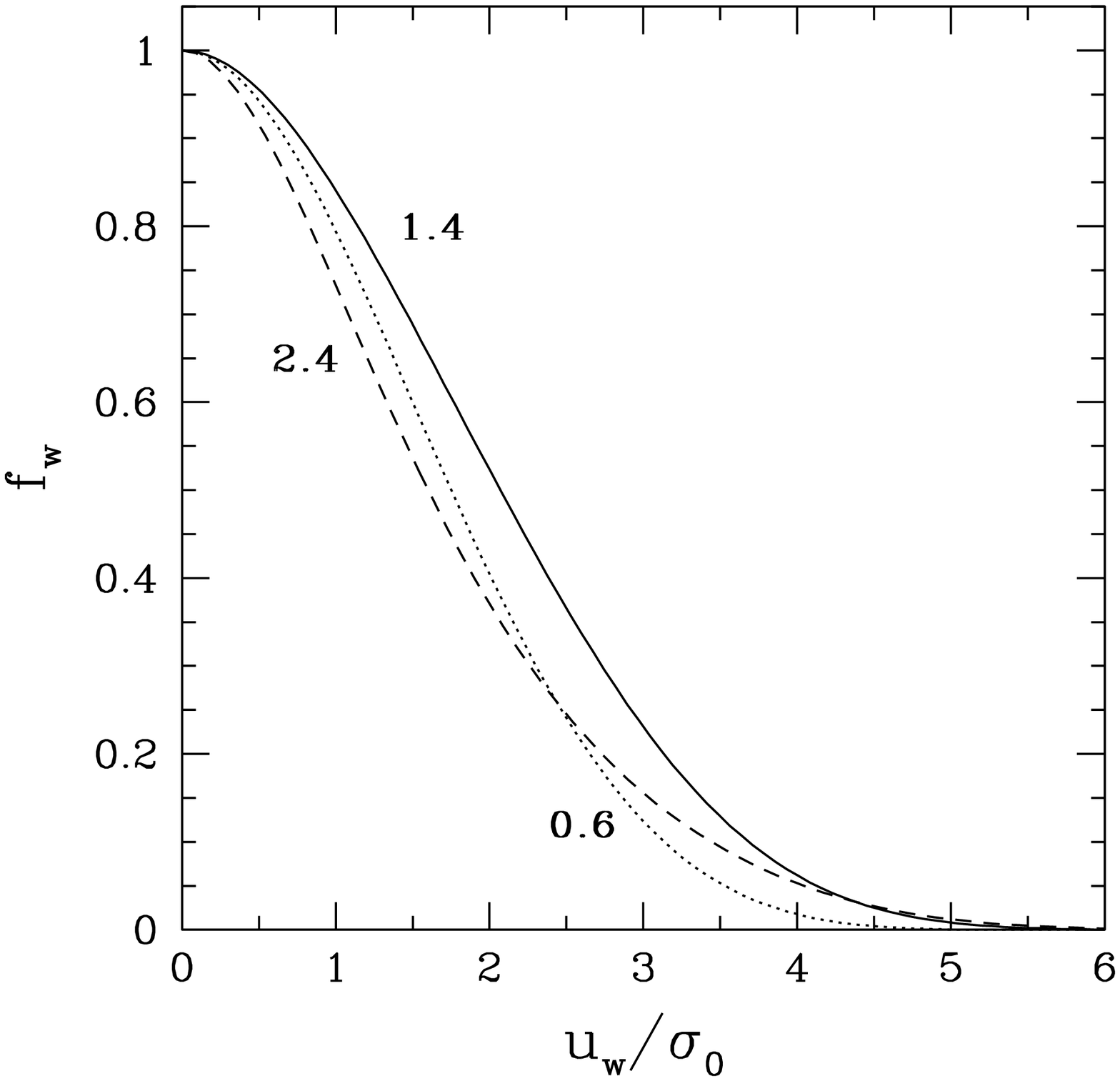,height=8.2cm}}
  \hspace*{0.5cm} Fig. 4.---{\small \cap4}
  \vspace*{0.5cm}
\else
  \begin{figure}
  \plotone{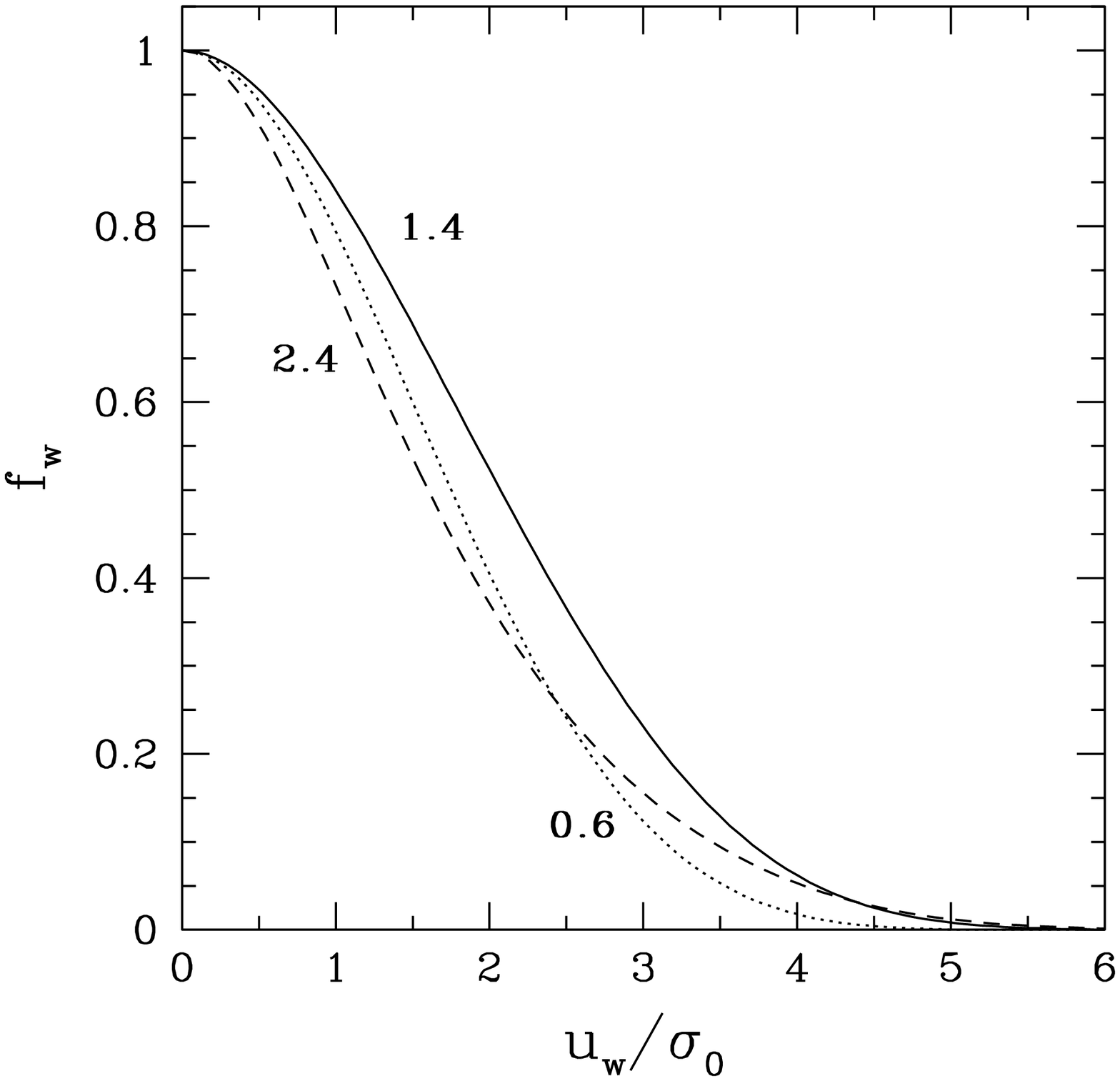} \label{fig:fw}
  \caption{\cap4}
  \end{figure}
\fi

\section{The orbit and the possibility of self-enrichment}

The question now is: could \omcen\ (and other clusters) retain the
wind material throughout their evolution in the Galaxy?  The oldest
stars in the Galactic disk are dated to be at least $9.5 \pm 0.8$ Gyr
old \citep{diskage1,diskage2}.  The age of the Universe is determined
fortuitously well from the coincidental CMB correlation \citep{KCS:01}
to be $14.0 \pm 0.5$ Gyr.  Thus, it is quite possible that the dense
gaseous disk was already assembled when the Universe was about 4 Gyr
old, equal to the age spread of stars in \omcen.  If this was the
case, ram pressure of the disk would strip any gas left in \omcen, as
we show below.

The current Galactocentric velocity of \omcen\ from the recently
measured proper motion \citep{Detal:99} is relatively low: $V_R = -31$
km s$^{-1}$, $V_\theta = -65$ km s$^{-1}$, $V_z = 4$ km s$^{-1}$, with
the uncertainty $\pm 10$ km s$^{-1}$.  The speed of only 70 km
s$^{-1}$ makes the cluster strongly bound and thus, its present
position at 6.3 kpc from the Galactic center is likely to be close to
the orbital apocenter.  Indeed, calculations of \citet{Detal:99} give
the pericenter distance $R_p = 1.2$ kpc and the apocenter distance
$R_a = 6.4$ kpc.  Therefore \omcen\ remains close enough to the
Galactic center to pass through the disk twice in each orbital period,
which is relatively short, $P_{\rm orb} = 1.2 \times 10^8$ yr.

We calculate the amount of gas that might have been present in \omcen\
as follows.  The total mass-loss rate of AGB stars depends on the
slope of the IMF between 1 and 8 $M_{\sun}$, $dN/dm \propto
m^{-\gamma}$, on the mass of the white dwarf remnant, $m_{wd}$, and on
the main sequence lifetimes of stars, $t_{ms}$:
\begin{equation}
{dM_w \over dt} = {dN \over dm} \, (m-m_{wd}) 
  \left(-{dt_{ms} \over dm}\right)^{-1},
\end{equation}
where $m = m_{ms}(t)$ is the inversion of the function $t_{ms}(m)$.
We consider two cases, $\gamma = 2.2$ (following \citealt{KTG:93}) and
$\gamma = 3.3$ \citep{S:86}.  Using the code of \citet{HPT:00}, we
calculate $t_{ms}(m)$ for the oldest stars in \omcen, with the
metallicity $[Fe/H]=-1.7$, and find that in the mass range of interest
it is well fit by $t_{ms} \approx 5\times 10^9\, (m/M_{\sun})^{-2.4}$
yr.  The masses of stellar remnants have been determined
observationally in the nearby star clusters \citep{W:00}, and the
result can be accurately fit by $m_{wd} = 0.444\, M_{\sun} + 0.084\,
m_{ms}$.  Combining all these expressions, we find the mass-loss rate
at the age of 4 Gyr: $dM_w/dt \approx 10^{-2}\, M_{cl}$ Gyr$^{-1}$.
Here $M_{cl} \approx 3\times 10^6\, M_{\sun}$ is the present mass of
the cluster.  (Interestingly, the mass-loss rate is almost the same
for both values $\gamma = 2.2$ and $\gamma = 3.3$.)  Thus, in half the
orbital period the cluster would accumulate $M_g = 2\times 10^3\,
M_{\sun}$ of gas.

We integrate the orbit of \omcen\ backward in time using an accurate
Galactic potential that fits most observational constraints of the
Milky Way (model $A_1$ of \citealt{KZS:02}).  We model the evolution
of the Galaxy starting with a spherical halo 14 Gyr ago, compress the
baryons into the disk and the bulge over an exponential timescale of 3
Gyr and then turn them into stars over the next 3 Gyr.  The bulge is
gradually converted into the present flattened shape.
%

We find the locations where the cluster crosses the plane of the disk
and then compare the ram pressure of the disk gas with the restoring
gravitational force per unit area at the half-mass radius of \omcen.
In many passages $\rho_{\rm disk} v_{\rm rel}^2 > G M_{cl} M_g / (4\pi
R_h^4)$ and therefore the AGB ejecta should be stripped from the
cluster.  Also, the current mass of \omcen\ is too small for dynamical
friction to affect the orbit.

\bigskip

To summarize, we have shown that if (1) \omcen\ had formed in
isolation with its present mass, and (2) the Galactic disk was in
place prior to the formation of the metal-rich stars, then the cluster
would encounter the disk gas in its orbital motion and lose the gas
accumulated from the enriched AGB winds.  
%
%
The same argument applies to all globular clusters within 10 kpc from
the Galactic center.  They could not have formed stars from the
material enriched by s-process elements.

Alternatively, if all the stars in the oldest globular clusters form
prior to the assembly of the Galactic disk, then \omcen\ must be
unrealistically special.  Why did it retain its enriched gas and
formed new populations of stars over 4 Gyr whereas a few dozen other
clusters, capable of doing so, did not.  As we have shown, even though
\omcen\ is the most massive cluster its potential well is similar to
many others.  So we are left again with the puzzle of why \omcen\ is
different from all other Galactic globular clusters.

{\it We conclude that it is highly unlikely that \omcen\ formed and
evolved in isolation and enriched itself with heavy elements of the
AGB winds.}

\section{In Closing: Towards a correct story of \omcen}

In light of the unusual properties of \omcen, it is becoming a popular
scenario that the cluster may be the core of a disrupted dwarf galaxy
(e.g., \citealt{F:93}).  The attraction is that a continuous infall of
gas to the center of the dwarf galaxy may lead to a variable star
formation history.  \citet{K:85} has shown that dwarf galaxies split
on the plot of the central surface brightness vs core radius into
dense dE galaxies (like M32) with the central velocity dispersion
$\sigma_0$ in excess of 50 km s$^{-1}$, and diffuse dSph galaxies
(like Draco) with $\sigma_0 < 20$ km s$^{-1}$.  Tidal stripping of the
outer envelope would decrease the total luminosity of the dwarf galaxy
but would not significantly change its core properties, $\mu_0$ and
$R_c$.  Thus, the cores of dE would have too high a velocity
dispersion and the cores of dSph would have too low density to count
as globular clusters.

The counter-examples to this rule are nucleated dSph galaxies,
sometimes classified as dE,N \citep{PC:93}.  NGC 205 \citep{M:01} has
a core velocity dispersion of 30 km s$^{-1}$, dropping to 15 km
s$^{-1}$ in the bright compact nucleus \citep{CS:90}.
%
%
The recently discovered Sagittarius galaxy apparently has a globular
cluster M54 positioned at its center \citep{IGI:94}.  Also,
\citet{Detal:00} detected five Ultra-Compact Dwarf galaxies in the
Fornax cluster of galaxies, which are brighter than globular clusters
but almost as compact.  They could be the result of mergers of young
star clusters, as suggested by \citet{FK:02}.  It remains to be seen
if the nuclei of these galaxies, when stripped of their envelopes,
would have similar kinematic and chemical properties to \omcen.  Also,
detailed orbital calculations are necessary to determine if the
cluster can be brought in to its present location near the Galactic
center (see \citealt{Z:01}).


The uniqueness of \omcen\ among the Galactic globular clusters still
poses unanswered questions.  In the hierarchical formation scenario we
expect a few dozen progenitor galaxies to make up the Milky Way.  If
\omcen\ did in fact form at the center of one of them, why did not
other almost as massive clusters form in the other progenitors and
have similar chemical enrichment?  Were they entirely disrupted or was
the history of \omcen\ completely different?  We intend to address
these issues in forthcoming work.

\acknowledgments 
We thank Chris Tout for discussions.  OG is supported by the STScI
Institute Fellowship, and JEP acknowledges the continuing support by
the STScI visitors program.  The electronic tables of the escape
velocities, velocity dispersions, and main-sequence lifetimes are
available at \mbox{\tt http://www.stsci.edu/$\sim$ognedin/gc/}.

\end{document}